 \documentclass[onecolumn,letterpaper,aps,prd,longbibliography,superscriptaddress,nofootinbib,floatfix]{revtex4-2}

\usepackage{graphicx}
\usepackage{xcolor}

\usepackage{xspace}	

\begin{document}


\title{
Cross sections of $\eta$ mesons in $p$$+$$p$ collisions at forward 
rapidity at $\sqrt{s}=500$ GeV and central rapidity at $\sqrt{s}=510$ GeV}

\newcommand{\abilene}{Abilene Christian University, Abilene, Texas 79699, USA}
\newcommand{\augie}{Department of Physics, Augustana University, Sioux Falls, South Dakota 57197, USA}
\newcommand{\banaras}{Department of Physics, Banaras Hindu University, Varanasi 221005, India}
\newcommand{\barc}{Bhabha Atomic Research Centre, Bombay 400 085, India}
\newcommand{\baruch}{Baruch College, City University of New York, New York, New York, 10010 USA}
\newcommand{\bnlcoll}{Collider-Accelerator Department, Brookhaven National Laboratory, Upton, New York 11973-5000, USA}
\newcommand{\bnlphys}{Physics Department, Brookhaven National Laboratory, Upton, New York 11973-5000, USA}
\newcommand{\caucr}{University of California-Riverside, Riverside, California 92521, USA}
\newcommand{\charlesczech}{Charles University, Faculty of Mathematics and Physics, 180 00 Troja, Prague, Czech Republic}
\newcommand{\ciae}{Science and Technology on Nuclear Data Laboratory, China Institute of Atomic Energy, Beijing 102413, People's Republic of China}
\newcommand{\cns}{Center for Nuclear Study, Graduate School of Science, University of Tokyo, 7-3-1 Hongo, Bunkyo, Tokyo 113-0033, Japan}
\newcommand{\colorado}{University of Colorado, Boulder, Colorado 80309, USA}
\newcommand{\columbia}{Columbia University, New York, New York 10027 and Nevis Laboratories, Irvington, New York 10533, USA}
\newcommand{\czechtech}{Czech Technical University, Zikova 4, 166 36 Prague 6, Czech Republic}
\newcommand{\dapnia}{Dapnia, CEA Saclay, F-91191, Gif-sur-Yvette, France}
\newcommand{\debrecen}{Debrecen University, H-4010 Debrecen, Egyetem t{\'e}r 1, Hungary}
\newcommand{\elte}{ELTE, E{\"o}tv{\"o}s Lor{\'a}nd University, H-1117 Budapest, P{\'a}zm{\'a}ny P.~s.~1/A, Hungary}
\newcommand{\ewha}{Ewha Womans University, Seoul 120-750, Korea}
\newcommand{\fsu}{Florida State University, Tallahassee, Florida 32306, USA}
\newcommand{\gsu}{Georgia State University, Atlanta, Georgia 30303, USA}
\newcommand{\hanyang}{Hanyang University, Seoul 133-792, Korea}
\newcommand{\hiroshima}{Physics Program and International Institute for Sustainability with Knotted Chiral Meta Matter (WPI-SKCM2), Hiroshima University, Higashi-Hiroshima, Hiroshima 739-8526, Japan}
\newcommand{\howard}{Department of Physics and Astronomy, Howard University, Washington, DC 20059, USA}
\newcommand{\hunrenatomki}{HUN-REN ATOMKI, H-4026 Debrecen, Bem t{\'e}r 18/c, Hungary}
\newcommand{\ihepprot}{IHEP Protvino, State Research Center of Russian Federation, Institute for High Energy Physics, Protvino, 142281, Russia}
\newcommand{\illuiuc}{University of Illinois at Urbana-Champaign, Urbana, Illinois 61801, USA}
\newcommand{\inrras}{Institute for Nuclear Research of the Russian Academy of Sciences, prospekt 60-letiya Oktyabrya 7a, Moscow 117312, Russia}
\newcommand{\instpasczech}{Institute of Physics, Academy of Sciences of the Czech Republic, Na Slovance 2, 182 21 Prague 8, Czech Republic}
\newcommand{\isu}{Iowa State University, Ames, Iowa 50011, USA}
\newcommand{\jaea}{Advanced Science Research Center, Japan Atomic Energy Agency, 2-4 Shirakata Shirane, Tokai-mura, Naka-gun, Ibaraki-ken 319-1195, Japan}
\newcommand{\jeonbuk}{Jeonbuk National University, Jeonju, 54896, Korea}
\newcommand{\jyvaskyla}{Helsinki Institute of Physics and University of Jyv{\"a}skyl{\"a}, P.O.Box 35, FI-40014 Jyv{\"a}skyl{\"a}, Finland}
\newcommand{\kek}{KEK, High Energy Accelerator Research Organization, Tsukuba, Ibaraki 305-0801, Japan}
\newcommand{\korea}{Korea University, Seoul 02841, Korea}
\newcommand{\kurchatov}{National Research Center ``Kurchatov Institute", Moscow, 123098 Russia}
\newcommand{\kyoto}{Kyoto University, Kyoto 606-8502, Japan}
\newcommand{\labllr}{Laboratoire Leprince-Ringuet, Ecole Polytechnique, CNRS-IN2P3, Route de Saclay, F-91128, Palaiseau, France}
\newcommand{\lahorelums}{Physics Department, Lahore University of Management Sciences, Lahore 54792, Pakistan}
\newcommand{\lawllnl}{Lawrence Livermore National Laboratory, Livermore, California 94550, USA}
\newcommand{\losalamos}{Los Alamos National Laboratory, Los Alamos, New Mexico 87545, USA}
\newcommand{\lpc}{LPC, Universit{\'e} Blaise Pascal, CNRS-IN2P3, Clermont-Fd, 63177 Aubiere Cedex, France}
\newcommand{\lund}{Department of Physics, Lund University, Box 118, SE-221 00 Lund, Sweden}
\newcommand{\lyon}{IPNL, CNRS/IN2P3, Univ Lyon, Universit{\'e} Lyon 1, F-69622, Villeurbanne, France}
\newcommand{\maryland}{University of Maryland, College Park, Maryland 20742, USA}
\newcommand{\mass}{Department of Physics, University of Massachusetts, Amherst, Massachusetts 01003-9337, USA}
\newcommand{\mate}{MATE, Institute of Technology, Laboratory of Femtoscopy, K\'aroly R\'obert Campus, H-3200 Gy\"ongy\"os, M\'atrai \'ut 36, Hungary}
\newcommand{\michigan}{Department of Physics, University of Michigan, Ann Arbor, Michigan 48109-1040, USA}
\newcommand{\miss}{Mississippi State University, Mississippi State, Mississippi 39762, USA}
\newcommand{\muenster}{Institut f\"ur Kernphysik, University of M\"unster, D-48149 M\"unster, Germany}
\newcommand{\muhlenberg}{Muhlenberg College, Allentown, Pennsylvania 18104-5586, USA}
\newcommand{\myongji}{Myongji University, Yongin, Kyonggido 449-728, Korea}
\newcommand{\nagasaki}{Nagasaki Institute of Applied Science, Nagasaki-shi, Nagasaki 851-0193, Japan}
\newcommand{\nara}{Nara Women's University, Kita-uoya Nishi-machi Nara 630-8506, Japan}
\newcommand{\natmephi}{National Research Nuclear University, MEPhI, Moscow Engineering Physics Institute, Moscow, 115409, Russia}
\newcommand{\newmex}{University of New Mexico, Albuquerque, New Mexico 87131, USA}
\newcommand{\nmsu}{New Mexico State University, Las Cruces, New Mexico 88003, USA}
\newcommand{\northcg}{Physics and Astronomy Department, University of North Carolina at Greensboro, Greensboro, North Carolina 27412, USA}
\newcommand{\ohio}{Department of Physics and Astronomy, Ohio University, Athens, Ohio 45701, USA}
\newcommand{\ornl}{Oak Ridge National Laboratory, Oak Ridge, Tennessee 37831, USA}
\newcommand{\orsay}{IPN-Orsay, Univ.~Paris-Sud, CNRS/IN2P3, Universit\'e Paris-Saclay, BP1, F-91406, Orsay, France}
\newcommand{\peking}{Peking University, Beijing 100871, People's Republic of China}
\newcommand{\pnpi}{PNPI, Petersburg Nuclear Physics Institute, Gatchina, Leningrad region, 188300, Russia}
\newcommand{\riken}{RIKEN Nishina Center for Accelerator-Based Science, Wako, Saitama 351-0198, Japan}
\newcommand{\rikjrbrc}{RIKEN BNL Research Center, Brookhaven National Laboratory, Upton, New York 11973-5000, USA}
\newcommand{\rikkyo}{Physics Department, Rikkyo University, 3-34-1 Nishi-Ikebukuro, Toshima, Tokyo 171-8501, Japan}
\newcommand{\saispbstu}{Saint Petersburg State Polytechnic University, St.~Petersburg, 195251 Russia}
\newcommand{\saopaulo}{Universidade de S{\~a}o Paulo, Instituto de F\'{\i}sica, Caixa Postal 66318, S{\~a}o Paulo CEP05315-970, Brazil}
\newcommand{\seoulnat}{Department of Physics and Astronomy, Seoul National University, Seoul 151-742, Korea}
\newcommand{\stonybrkc}{Chemistry Department, Stony Brook University, SUNY, Stony Brook, New York 11794-3400, USA}
\newcommand{\stonycrkp}{Department of Physics and Astronomy, Stony Brook University, SUNY, Stony Brook, New York 11794-3800, USA}
\newcommand{\sungskku}{Sungkyunkwan University, Suwon, 440-746, Korea}
\newcommand{\tenn}{University of Tennessee, Knoxville, Tennessee 37996, USA}
\newcommand{\titech}{Department of Physics, Tokyo Institute of Technology, Oh-okayama, Meguro, Tokyo 152-8551, Japan}
\newcommand{\tsukuba}{Tomonaga Center for the History of the Universe, University of Tsukuba, Tsukuba, Ibaraki 305, Japan}
\newcommand{\usmma}{United States Merchant Marine Academy, Kings Point, New York 11024, USA}
\newcommand{\vandy}{Vanderbilt University, Nashville, Tennessee 37235, USA}
\newcommand{\weizmann}{Weizmann Institute, Rehovot 76100, Israel}
\newcommand{\wigner}{Institute for Particle and Nuclear Physics, HUN-REN Wigner Research Centre for Physics, (HUN-REN Wigner RCP, RMI), H-1525 Budapest 114, POBox 49, Budapest, Hungary}
\newcommand{\yonsei}{Yonsei University, IPAP, Seoul 120-749, Korea}
\newcommand{\zagreb}{Department of Physics, Faculty of Science, University of Zagreb, Bijeni\v{c}ka c.~32 HR-10002 Zagreb, Croatia}
\newcommand{\zambia}{Department of Physics, School of Natural Sciences, University of Zambia, Great East Road Campus, Box 32379, Lusaka, Zambia}
\affiliation{\abilene}
\affiliation{\augie}
\affiliation{\banaras}
\affiliation{\barc}
\affiliation{\baruch}
\affiliation{\bnlcoll}
\affiliation{\bnlphys}
\affiliation{\caucr}
\affiliation{\charlesczech}
\affiliation{\ciae}
\affiliation{\cns}
\affiliation{\colorado}
\affiliation{\columbia}
\affiliation{\czechtech}
\affiliation{\dapnia}
\affiliation{\debrecen}
\affiliation{\elte}
\affiliation{\ewha}
\affiliation{\fsu}
\affiliation{\gsu}
\affiliation{\hanyang}
\affiliation{\hiroshima}
\affiliation{\howard}
\affiliation{\hunrenatomki}
\affiliation{\ihepprot}
\affiliation{\illuiuc}
\affiliation{\inrras}
\affiliation{\instpasczech}
\affiliation{\isu}
\affiliation{\jaea}
\affiliation{\jeonbuk}
\affiliation{\jyvaskyla}
\affiliation{\kek}
\affiliation{\korea}
\affiliation{\kurchatov}
\affiliation{\kyoto}
\affiliation{\labllr}
\affiliation{\lahorelums}
\affiliation{\lawllnl}
\affiliation{\losalamos}
\affiliation{\lpc}
\affiliation{\lund}
\affiliation{\lyon}
\affiliation{\maryland}
\affiliation{\mass}
\affiliation{\mate}
\affiliation{\michigan}
\affiliation{\miss}
\affiliation{\muenster}
\affiliation{\muhlenberg}
\affiliation{\myongji}
\affiliation{\nagasaki}
\affiliation{\nara}
\affiliation{\natmephi}
\affiliation{\newmex}
\affiliation{\nmsu}
\affiliation{\northcg}
\affiliation{\ohio}
\affiliation{\ornl}
\affiliation{\orsay}
\affiliation{\peking}
\affiliation{\pnpi}
\affiliation{\riken}
\affiliation{\rikjrbrc}
\affiliation{\rikkyo}
\affiliation{\saispbstu}
\affiliation{\saopaulo}
\affiliation{\seoulnat}
\affiliation{\stonybrkc}
\affiliation{\stonycrkp}
\affiliation{\sungskku}
\affiliation{\tenn}
\affiliation{\titech}
\affiliation{\tsukuba}
\affiliation{\usmma}
\affiliation{\vandy}
\affiliation{\weizmann}
\affiliation{\wigner}
\affiliation{\yonsei}
\affiliation{\zagreb}
\affiliation{\zambia}
\author{N.J.~Abdulameer} \affiliation{\debrecen} \affiliation{\hunrenatomki}
\author{U.~Acharya} \affiliation{\gsu}
\author{A.~Adare} \affiliation{\colorado} 
\author{C.~Aidala} \affiliation{\losalamos} \affiliation{\michigan} 
\author{N.N.~Ajitanand} \altaffiliation{Deceased} \affiliation{\stonybrkc} 
\author{Y.~Akiba} \email[PHENIX Spokesperson: ]{akiba@rcf.rhic.bnl.gov} \affiliation{\riken} \affiliation{\rikjrbrc}
\author{R.~Akimoto} \affiliation{\cns} 
\author{H.~Al-Ta'ani} \affiliation{\nmsu} 
\author{J.~Alexander} \affiliation{\stonybrkc} 
\author{M.~Alfred} \affiliation{\howard} 
% ...
\collaboration{PHENIX Collaboration}  \noaffiliation

\date{\today}


\begin{abstract}


We present the first measurements of the forward and midrapidity 
$\eta$-meson cross sections from $p$$+$$p$ collisions at $\sqrt{s}=500$ 
and $510$~GeV, respectively. We also report the midrapidity 
$\eta/\pi^0$ ratio at 510 GeV. The forward cross section is measured 
differentially in $\eta$-meson transverse momentum ($p_T$) from 1.0 to 
6.5~GeV/$c$ for pseudorapidity $3.0<|\eta|<3.8$. The midrapidity cross 
section is measured from 3.5 to 44 GeV/$c$ for pseudorapidity 
$|\eta|<0.35$. Both cross sections serve as critical inputs to an 
updated global analysis of the $\eta$-meson fragmentation functions.

\end{abstract}

\maketitle

\section{\label{sec:introduction}Introduction}

Experimental measurements from hadronic collisions have served as an 
invaluable tool in the development of quantum chromodynamics (QCD). 
While quarks and gluons, together known as partons, are the fundamental 
particles of QCD, the experimentally relevant degrees of freedom are 
their hadronic bound states. The relationship between the partons and 
hadrons is contained in nonperturbative functions that rely on 
experimental input for their description. Take, for example, the 
inclusive production of hadrons from proton collisions, 
$p+p{\rightarrow}h+X$. At short distance scales, the partonic hard 
scattering matrix elements are calculable in fixed-order perturbative 
QCD (pQCD). However, the long-distance initial- and final-state 
descriptions are nonperturbative. The initial state is described by the 
parton distribution functions (PDFs) $f(x,Q^2)$ which, to leading order 
in $\alpha_s$, give the number density of a parton in the proton at a 
given collinear momentum fraction $x$ and squared momentum transfer 
scale $Q^2$. In the final state, the scattered partons undergo 
hadronization, where they evolve from the hard scattering to the 
observable hadronic-QCD bound states. This process is described by 
fragmentation functions (FFs) $D(z,Q^2)$, which, to leading order, give 
the probability density that a struck parton will fragment into a given 
final state hadron with collinear momentum fraction $z$ at $Q^2$.
QCD factorization enables the separation of the nonperturbative 
functions from the perturbative physics~\cite{Collins:1989gx}. 
Hadronic-level observables can then be calculated by convoluting the 
initial-state PDFs, final-state FFs, and the partonic scattering cross 
section described by pQCD. The PDFs and FFs, which are usually assumed 
to be universal, are determined through global fits to an array of 
experimental results.

Fragmentation into the lightest QCD bound state, the pion, has been 
thoroughly studied with spectra of $\pi^0$ and $\pi^{\pm}$ from 
hadron-hadron 
collisions~\cite{ADAMS2006161,PhysRevLett.97.152302,PhysRevD.80.111108,PhysRevLett.108.072302, 
PhysRevD.89.012001,PhysRevD.76.051106,PhysRevD.93.011501,ALICE:2012wos,2018EPJC.78.263A,EPJ.C77.339,ALICE_2021},
$\pi^{\pm}$ from electron-positron 
annihilation~\cite{PhysRevD.59.052001,Z.Phys.C.66,refId0,PhysRevD.101.092004,PhysRevD.88.032011} 
and $\pi^{\pm}$ from semiinclusive deep-inelastic 
scattering (SIDIS)~\cite{PhysRevD.87.074029,ADOLPH20171}.
Global analyses of these results have constrained the pion FFs to high 
precision~\cite{PhysRevD.75.114010, PhysRevD.91.014035, 
PhysRevD.105.L031502}. In contrast, the AESSS~(2011)~\cite{PhysRevD.83.034002} 
$\eta$-meson FFs from over a decade ago are the one global analysis 
available for $\eta$ mesons and are significantly less constrained. In the 
time since these FFs were initially determined, more measurements of $\eta$ meson 
production have been made over a wide range of energies in hadronic 
collisions~\cite{PhysRevD.90.072008,PhysRevC.81.064904,ALICE:2012wos,PhysRevC.109.024907,2018EPJC.78.263A,EPJ.C77.339} 
and in electron-positron annihilation~\cite{PhysRevD.111.052003}. Large 
deviations from predictions using the AESSS~FFs have been found in the 
high-energy data at the Large Hadron Collider (LHC), motivating an update to 
the $\eta$ meson FF global fits. The measurements of the $\eta$-meson cross 
sections in the forward region at $\sqrt{s}=500$ GeV and central region at 
$\sqrt{s}=510$ GeV presented here serve as two principal inputs 
in a newly released global analysis of the $\eta$ meson FFs, denoted the 
ALMSS~(2025)~FFs~\cite{etaFFs}. The cross sections are the first of their kind at 
$\sqrt{s}=500$ GeV and $\sqrt{s}=510$~GeV, providing an important 
intermediate energy input into the ALMSS~FFs between the lower-energy data 
at the Relativistic Heavy Ion Collider (RHIC) and higher-energy data at the 
LHC. The cross sections are compared to next-to-leading-order (NLO) pQCD 
predictions utilizing the earlier 
AESSS~(2011)~FFs~\cite{PhysRevD.83.034002,PhysRevD.67.054005,Vogelsang_priv} 
and the newer ALMSS~(2025)~FFs~\cite{etaFFs}.

\section{\label{sec:experiment}The PHENIX experiment at RHIC}

\subsection{\label{sec:phenix}PHENIX detector}

\begin{figure}[htb]
\includegraphics[width=1.0\linewidth]{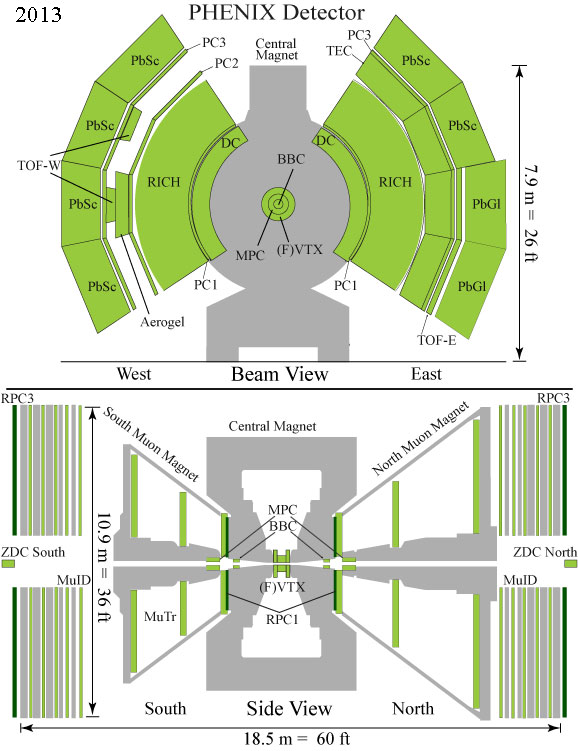}
\caption{\label{fig:phenix} 
The PHENIX detector during the 2013 RHIC 
run.}
\end{figure}

The PHENIX experiment~\cite{ADCOX2003469}, shown in Fig.~\ref{fig:phenix}, 
was located in the 8 o'clock interaction region of RHIC at Brookhaven 
National Laboratory and operated from 2000-2016. The detector comprised two 
central-arm spectrometers covering pseudorapidity $|\eta|<0.35$, two muon 
arms covering $1.2<|\eta|<2.4$, two global detectors, which are the 
beam-beam counters (BBC) and zero-degree calorimeters, plus a forward 
electromagnetic calorimeter covering $3.1<|\eta|<3.9$. 

The central-arm spectrometers each covered $\pi/2$ in azimuth and comprised 
($i$) a high-granularity electromagnetic calorimeter 
(EMCal)~\cite{APHECETCHE2003521}, ($ii$) drift and pad chambers for charged-track 
reconstruction~\cite{ADCOX2003489}, and ($iii$) a ring-imaging \v{C}erenkov detector 
for charged-particle identification~\cite{AIZAWA2003508}. The muon 
arms~\cite{AKIKAWA2003537} had a combination of tracking and identification 
capabilities and were used primarily for detecting vector mesons through 
their dimuon decays. The BBC~\cite{ALLEN2003549} was an 
array of quartz \v{C}erenkov radiators covering $3.0<|\eta|<3.9$ used for 
minimum-bias (MB) triggering on inelastic proton collisions. It also served 
to measure luminosity and determine the event vertex position along the beam 
axis (z~vertex) to ${\pm}2$~cm.

\subsection{\label{sec:mpc}Muon-Piston Calorimeter}

The muon-piston calorimeter (MPC) was a forward electromagnetic 
calorimeter primarily used for measurements of $\pi^0$ and $\eta$ 
mesons through their diphoton decay. The MPC had two arms with full 
azimuthal acceptance that were each located $\pm$ 220~cm away from the 
nominal PHENIX interaction point. The south MPC covered 
$-3.8<\eta<-3.1$ and the north MPC covered $3.1<\eta<3.9$. The south 
(north) MPC comprised 196 (220) $2.2\times2.2\times18$~cm$^3$ 
$\textrm{PbWO}_4$ crystal towers read out by Hamamatsu S8664-55 
avalanche photodiodes. Tower-by-tower gain variations in time were 
tracked and corrected by an LED calibration system. An absolute 
calibration was performed using an iterative algorithm to match the 
$\pi^0$ invariant-mass peak position of diphoton clusters. The nominal 
relative energy resolution of the MPC was 
$\sigma(E)/E=13\%/\sqrt{E\rm [GeV]}\oplus{8}$\% 
with an energy scale uncertainty of 
2\%~\cite{PhysRevD.90.072008,PhysRevD.90.012006}. A dedicated MPC trigger 
was utilized for high-$p_T$ $\eta$ mesons that required the energy sum of a 
4-by-4 group of towers (tile) to exceed a certain threshold, which at 
$\sqrt{s}=500$ GeV corresponds to approximately 40 GeV. The set of 56 (61) 
4-by-4 tiles in the south (north) MPC are in general nondisjoint, 
overlapping by up to two towers in both the horizontal and vertical 
directions to reduce the impact of a position-dependent threshold tied to 
cluster showers that can spread their energy across multiple tiles.

\subsection{\label{sec:emcal}Central Electromagnetic Calorimeter}

The PHENIX central EMCal provided highly 
segmented detection of photons and electrons in two nearly back-to back 
arms each covering a pseudorapidity of $|\eta|<0.35$ and 
$\phi\approx\pi/2$ in azimuth. The EMCal comprised 8 sectors with 24,768 
calorimeter towers of two different calorimeter technologies, with 75\% 
of the acceptance (six sectors; four in the west arm and two in the 
east arm) covered by a lead scintillator (PbSc) sampling calorimeter 
and the remaining 25\% (two sectors in the east arm) covered by a lead 
glass (PbGl) \v{C}erenkov calorimeter. The EMCal granularity was 
$\Delta\phi \times\Delta\eta=0.011\times 
0.011\left(0.008\times{0.008}\right)$ and the nominal relative-energy 
resolution for the PbSc (PbGl) calorimeters~\cite{APHECETCHE2003521}
was $\sigma(E)/E=8.1\%/\sqrt{E\rm [GeV]}\oplus 2\%$($5.9\% / 
\sqrt{E\rm [GeV]}\oplus 0.8\%$).  The EMCal also provided particle 
time-of-flight (ToF) measurements, an essential tool for reducing the 
impact of out-of-time pileup that could accrue over integration 
time. The energy calibration of the EMCal was performed 
by matching the reconstructed $\pi^0$ mass for each calorimeter tower. 
The ToF calibration was performed on a run-by-run basis by fitting to 
the peak of the photon time-of-flight distribution on a given tower. A 
dedicated EMCal high-momentum photon (HPP) trigger, similar to the MPC 
4-by-4 trigger, required the energy deposition in a group of 4-by-4 
towers to exceed a threshold corresponding to approximately 4 GeV for 
$\sqrt{s}=510$ GeV collisions. This trigger enabled the measurement 
of high-$p_T$ mesons and contained information on which photon clusters 
and calorimeter sectors in an event fired the trigger.

\section{\label{sec:etaxs}The $\eta$-meson cross section}

The invariant cross section of inclusive $\eta$ meson production
$p+p\rightarrow\eta+X$ was measured by 
\begin{equation} \label{eq:1} 
E\frac{d^3\sigma}{dp^3} = 
\frac{1}{\mathcal{L}}\frac{1}{\mathcal{{\rm BR}_{\eta\rightarrow 
\gamma\gamma}}}\frac{1}{2\pi p_T}\frac{\Delta N^{\rm meas}}{\epsilon_{\rm trig} \epsilon_{\rm reco} \Delta p_T \Delta \eta}, 
\end{equation} 

\noindent where $\Delta N^{\rm meas}$ is the measured $\eta$ meson 
yield in a given bin of $p_T$ and pseudorapidity, $\epsilon_{\rm trig}$ 
is the efficiency with which the MB or dedicated high $p_T$ trigger 
will fire when an $\eta$ meson is produced, $\epsilon_{\rm reco}$ is the 
efficiency of the detector to reconstruct an $\eta$ meson through its 
decay to diphotons, $\mathcal{L}$ is the integrated luminosity of the 
dataset, and $\mathcal{{\rm BR}_{\eta\rightarrow\gamma\gamma}}=0.3936$ is 
the branching ratio (fraction) of the 
$\eta\rightarrow\gamma\gamma$~decay~\cite{PhysRevD.110.030001}.

The integrated luminosity is determined as the number of sampled events in 
the dataset divided by the cross section of inelastic $p$$+$$p$ collisions 
that the BBCs detected. The BBC cross section at $\sqrt{s}=500$ GeV was 32.5 
mb with a 9.3\% systematic uncertainty, as measured by a Vernier scan of 
events within $\pm 30$ cm from the nominal interaction 
point~\cite{PhysRevD.93.011501}.

For the forward $\eta$-meson cross section, a subpercent 
level correction was applied to account for the more 
lenient BBC vertex cut of $\pm70$~cm, leading to a final 
BBC cross section of 32.25 mb. The integrated luminosities of the 
MB and MPC-triggered datasets used in the analysis were 
$\mathcal{L}_{\rm MB}=5.56\times10^{-3}$ pb$^{-1}$ and 
$\mathcal{L}_{\rm MPC}=9.33$ pb$^{-1}$. Any multiple-collision effects 
on the luminosity counting were found to be negligible with the 
relatively small number of interactions per BBC triggered event, 
$\left<n_{\rm coll}\right>{\approx}1.1$.

For the midrapidity $\eta$ meson cross section, the integrated luminosities 
of the MB and HPP trigger datasets used in the analysis were 
$\mathcal{L}_{\rm MB}=8.45\times10^{-3}$ pb$^{-1}$ and 
$\mathcal{L}_{\rm HPP}=13.77$ pb$^{-1}$. 
Large instantaneous luminosities meant 
that up to one third of all bunch crossings contained more than one 
$p$$+$$p$ collision. Luminosity corrections for multiple collisions of up 
to 15\% were calculated independently for the PbSc and PbGl calorimeters in 
MB and HPP datasets. This was done by studying the behavior of the ratio of 
the $\pi^0$ yield to the number of MB triggers as a function of the 
instantaneous MB trigger rate~\cite{PhysRevD.93.011501}.

\subsection{\label{ssec:etaxs}The forward $\eta$-meson cross section 
at $\sqrt{s}=500$~GeV} 

In 2009, PHENIX collected data from proton-proton collisions at 
$\sqrt{s}=500$ GeV. The cross section was measured independently in the south 
and north MPCs in MB and MPC-triggered datasets. In events that satisfied 
the MB requirement, the cross section was measured differentially in $p_T$ 
from 1.0 to 4.5 GeV/$c$. Events that fired the MPC trigger provided a high 
$p_T$ sample from 3.5 to 6.5 GeV/$c$.

\subsubsection{\label{sssec:forwardetayields}Yields}

\begin{figure*}[htb]
\includegraphics[width=0.99\linewidth]{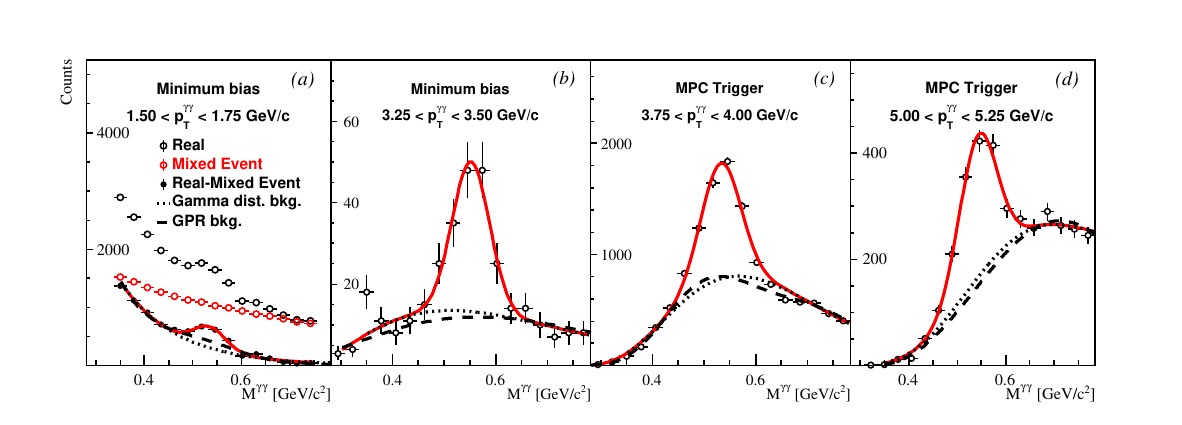}
\caption{\label{fig:invmass}Examples of two-photon invariant-mass 
distributions in the MPC from the 2009 $\sqrt{s}=500$ GeV run in 
MB [(a),(b)] and MPC-triggered data [(c),(d)]. The black open 
circles represent the real diphoton cluster pairs, the red open circles 
represent diphoton cluster pairs from the mixed-event background, and 
the black closed circles represent the real cluster pairs with the 
mixed-event background subtracted. The red solid curves show the 
combined fit of the gamma-distribution background and signal while the gray dotted curves 
show only the gamma-distribution background fit. The black dashed curves show the background description from Gaussian process regression.}
\end{figure*}

The raw yields of $\eta$ mesons are extracted from the invariant-mass 
distributions of photon cluster pairs as shown in 
Fig.~\ref{fig:invmass}. The invariant mass is given by

\begin{equation} \label{eq:2}
M_{\gamma \gamma} = \sqrt{E_{\gamma \gamma}^2 - \vec{p}_{\gamma \gamma} \cdot \vec{p}_{\gamma \gamma}} = \sqrt{4E_1E_2}\sin{\frac{\theta}{2}},
\end{equation}
where the $E_1,E_2$ are the cluster energies and $\theta$ is the 
opening angle between the cluster momentum vectors. A cut on the 
cluster energy asymmetry $\alpha=|E_2-E_1|/(E_2+E_1)<0.6$ is used to 
reduce the background from mismatched photon clusters. Additionally, 
cluster pairs are required to be separated by at least 2.6 cm in the 
MPC to lessen the impact of high-momentum $\eta$ mesons whose photon daughters have an opening angle small enough to be reconstructed in the same cluster, known as a merged cluster.

For cluster-pair transverse momentum $p_T^{\gamma \gamma}$ below 
2.25~GeV/$c$ in the MB dataset, a mixed-event background 
subtraction is used to reject pairs from combinatorial background 
sources. The mixed-event distribution is formed by pairing clusters 
from distinct events with one another, removing any likelihood of 
creating correlated pairs. Above 2.25~GeV/$c$, the impact of any mixed 
event background subtraction is minimal, so yields were determined by 
directly fitting to the real pairs.

The raw $\eta$ meson yields are determined using two independent fitting 
methods. First, the background with the signal region removed 
($0.45<M_{\gamma \gamma}<0.65$~GeV/$c^2$) is fit to a gamma-distribution 
function. In the case of the MPC-triggered data, the functional form of the 
background also includes a Gaussian centered at 0.78~GeV/$c^2$ to describe a 
small resonance from $\omega(782) \rightarrow \pi^0 + \gamma$ decays where 
the $\pi^0$ has decayed into two photons which merge into a single cluster. 
The signal region is then included back into the invariant-mass distribution 
and a simultaneous fit of the background fixed with the previously 
determined parameters and the signal, described by a Gaussian, was used to 
extract the raw yields. The second method fitted the background distributions 
with the signal region removed using Gaussian process regression 
(GPR)~\cite{Rasmussen2006} with a radial-basis-function kernel. The yields 
are extracted as the difference of integrals in the signal region of the 
total invariant-mass distribution and the GPR fit of the background. The 
final raw yields in Fig.~\ref{fig:yields_trigeff_recoeff_forward}(a) are the 
weighted mean of those found between the two methods, and a systematic 
uncertainty is taken as their relative difference. The yield extraction was 
found to be stable across small changes in the mass fit region.

\subsubsection{\label{sssec:forwardeffs}Efficiencies and Corrections}

\begin{figure*}[htb]
\includegraphics[width=0.99\linewidth]{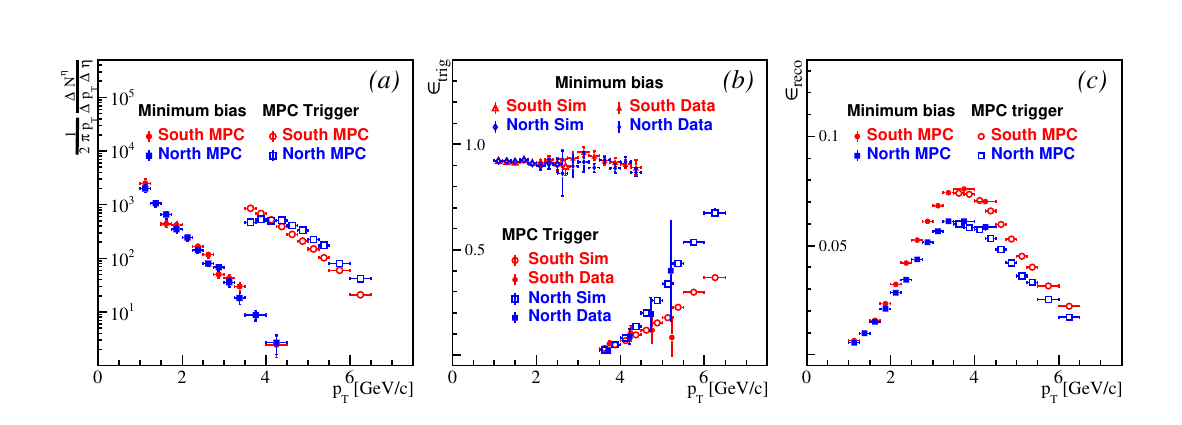}
\caption{\label{fig:yields_trigeff_recoeff_forward}The inputs to the forward 
$\eta$ meson cross section at $\sqrt{s}=500$ GeV for both the MB and 
MPC-triggered dataset at pseudorapidities $3.0<|\eta|<3.8$. (a) The raw 
yields as a function of $\eta$ meson $p_T$, (b) the trigger efficiency as a 
function of $\eta$ meson $p_T$, and (c) the reconstruction efficiency as a 
function of $\eta$ meson $p_T$.}
\end{figure*}

\begin{figure}[htb]
\includegraphics[width=1.0\linewidth]{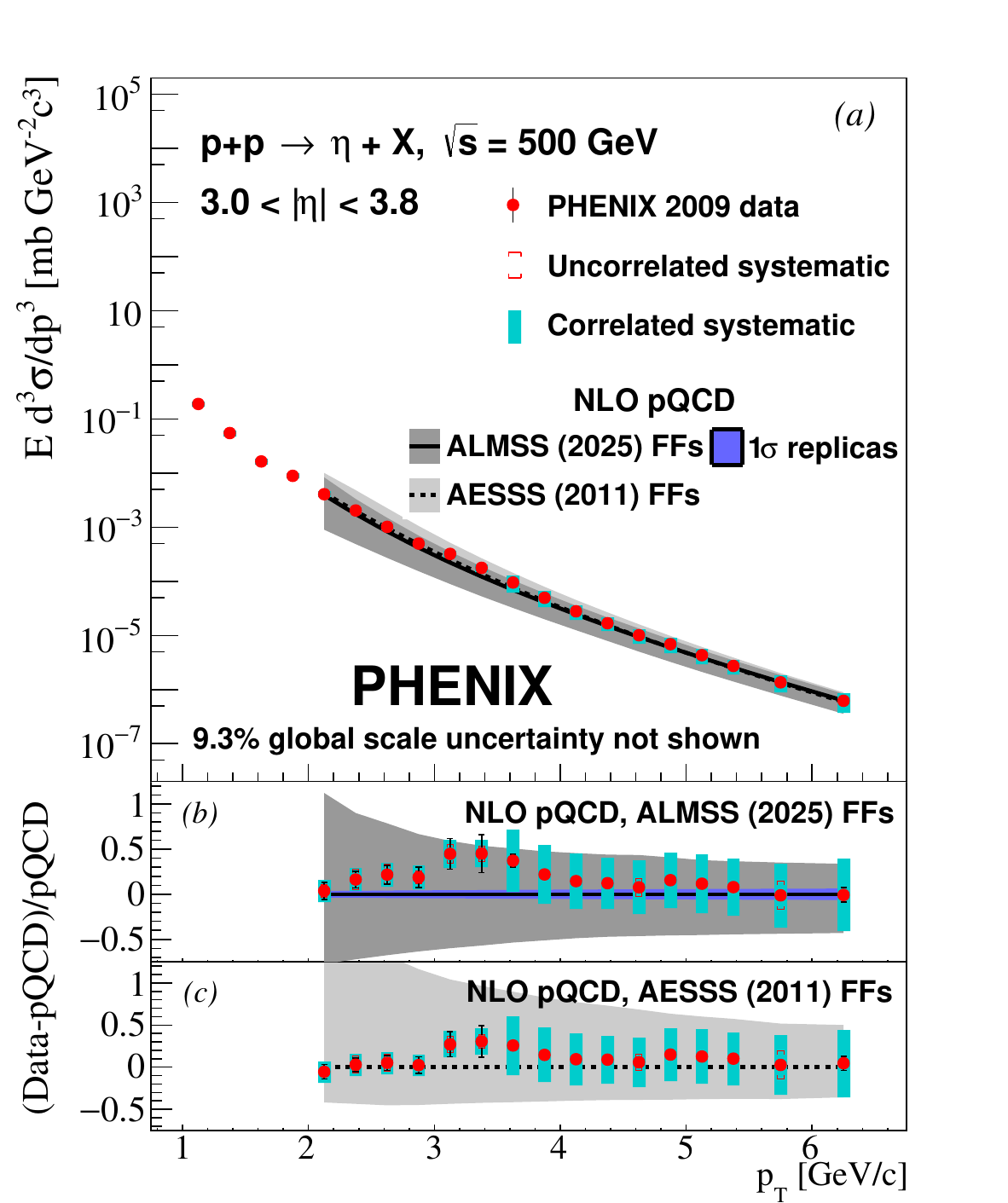}
\caption{\label{fig:finalcrosssection}The invariant cross section of 
inclusive $\eta$ mesons at forward rapidity in $\sqrt{s}=500$ GeV $p$$+$$p$ 
collisions. Statistical uncertainties are shown in black uncertainty bars, 
systematic uncertainties uncorrelated across $p_T$ are shown in red 
uncertainty bars, and systematic uncertainties correlated across $p_T$ are 
shown in blue uncertainty boxes. A global 9.3\% systematic uncertainty is 
not shown. NLO-pQCD predictions using the AESSS~(2011)~\cite{PhysRevD.83.034002} 
$\eta$~FFs and the ALMSS~(2025)~\cite{etaFFs} $\eta$~FFs are shown as the dashed 
and solid line, respectively, and their corresponding theoretical scale 
uncertainties are displayed as the gray uncertainty bands. The blue 
uncertainty band shows the 1$\sigma$ standard deviation of the cross section 
predictions across the final set of ALMSS $\eta$ FF replicas.} 
\end{figure}

The raw yields must be corrected by the fraction of produced $\eta$ mesons 
that are detected by the MPC. This fraction was encoded in two efficiencies: 
a trigger efficiency, which represents the probability that an event with an 
$\eta$ meson was successfully triggered, and the reconstruction efficiency, 
which represents the probability that the produced $\eta$ meson will be in 
the acceptance of the MPC and could have been reconstructed as a diphoton 
decay within the $\eta$ meson signal region, accounting for smearing effects 
due to the finite energy and spatial resolution of the detector and the 
selection requirements used to identify diphoton candidates. The 
reconstruction efficiency also accounted for any cluster-merging effects.

To help calculate these corrections, a set of simulated single $\eta$ mesons 
was generated in Monte Carlo as a flat distribution in $p_T$ from 0.5 
to 8.5 GeV/$c$ and pseudorapidity for $2.5<|\eta|<4.5$, passed 
through a {\sc geant3}~\cite{AGOSTINELLI2003250} description of the PHENIX 
detector, and embedded into real MB events. Photon cluster 
pairs are identified from these simulated $\eta$ mesons with the same 
reconstruction work flow and cluster kinematic cuts as real data. The 
simulations are weighted to match the real event z-vertex distribution 
and the variations of tower gain over time.  

For $p_T>2.5$ GeV/$c$, the MB trigger efficiency was calculated in a data-driven 
manner by finding the fraction of $\eta$ mesons that fire the 
MB trigger in the MPC-triggered dataset. This value was found to be 
uniform in $p_T$ at $0.92 \pm 0.01\textrm{(stat)} \pm 
0.04\textrm{(syst)}$ where the systematic uncertainty is assigned for 
the small difference between the constant values found in the south and 
north MPCs, as seen in Fig.~\ref{fig:yields_trigeff_recoeff_forward}(b). This 
efficiency agrees with a previous PHENIX midrapidity $\pi^0$ analysis 
at 510 GeV~\cite{PhysRevD.93.011501}, which measured the efficiency 
down to 1 GeV/$c$. We confirmed this uniform behavior of the efficiency 
at low $p_T<2.5$ GeV/$c$ by finding the fraction of charged particles 
from {\sc pythia6} 
(TuneA)~\cite{Pythia6,field2002underlyingeventhardscattering,TeV4LHCQCDWorkingGroup:2006fht}
simulations that fall into the BBC acceptance.

Insufficient statistics at high $p_T$ in the MB dataset 
prevent a fully data-driven estimation of the MPC trigger efficiency. 
Instead, a technique for simulating the trigger was 
utilized~\cite{PhysRevD.90.072008}. First, trigger efficiencies as a 
function of raw ADC were gathered for each MPC 4-by-4 tile from data 
throughout the entire data-taking period. At any given time, these 
efficiencies are a step function of the ADC, but variations in gain 
over time spread the step functions to some extent. The 
time-integrated tile efficiencies were then fitted by a 
sigmoid-like double error function: 
\begin{equation} \label{eq:3} f(x) = 
\int_{-\infty}^x \left[a\mathcal{N}_1(x';\mu_1,\sigma_1) + 
b\mathcal{N}_2(x';\mu_2,\sigma_2)\right]dx', 
\end{equation}

\noindent where $\mathcal{N}_{1,2}$ are Gaussian distributions. To find 
the ADC thresholds ($ADC_{\rm thresh}$) that correspond to the turn-on 
value of the step functions associated with each tile, candidate values 
of $\theta_{\rm thresh}=f(ADC_{\rm thresh})$ were tested by matching single 
cluster trigger efficiencies from full {\sc pythia6} (TuneA) + 
{\sc geant3} simulations and data. A simulated cluster is considered to have 
fired the trigger if at least one of its associated tile ADCs exceeds 
the ADC value that maps to the chosen $\theta_{\rm thresh}$ determined by 
the tile fit parameters from Eq.~(\ref{eq:3}). The optimal threshold 
for matching the simulation to the data was determined to be 
$\theta_{\rm thresh}=0.73$ which for 500 GeV $p$$+$$p$ collisions in 2009 
corresponds to the 4-by-4 tile thresholds of $\approx$ 40 GeV. A 
trigger efficiency systematic was taken for slight differences in the 
threshold determination between the south and north MPCs.  The final 
MPC trigger efficiency binned in $\eta$ meson $p_T$ was found by 
examining whether any of the tiles associated with either of the 
reconstructed photon clusters from the single $\eta$ meson simulations 
exceeded the optimal ADC threshold. At low $p_T$ where MB data can be 
used, the resulting trigger efficiency shown in 
Fig.~\ref{fig:yields_trigeff_recoeff_forward}(b) agrees with that of 
real photon clusters from the data.

The calculation of the reconstruction efficiency relies on an iterative 
weighting procedure of the simulated $\eta$ meson sample to replicate 
realistic $p_T$ and pseudorapidity distributions. On an initial pass, 
the reconstruction efficiency was calculated as the fraction of 
reconstructed to generated $\eta$ mesons where each $\eta$ meson has 
been weighted for its $p_T$ and pseudorapidity from {\sc pythia6} (TuneA) 
truth spectra. This weighted efficiency is then used as a correction to 
measured yields from real data. The resulting corrected real spectra 
are used as the weights to the efficiency calculation of the next 
iteration. After several iterations, the reconstruction efficiencies 
converge to the stable result shown in 
Fig.~\ref{fig:yields_trigeff_recoeff_forward}(c). The decrease in reconstruction 
efficiency at $p_T \gtrsim$ 4 GeV/$c$ stems from enhanced cluster merging 
of the diphoton clusters from boosted $\eta$ mesons. Differences 
between the south and north MPC reconstruction efficiencies are due to 
the larger number of towers in the north MPC that are removed in 
quality assurance due to an anomalous driver board and beam pipe 
support blockage.

\subsubsection{\label{sec:systs}Systematic Uncertainties} 

Systematic uncertainties for the forward cross section are determined 
independently for the south and north arms of the MPC, and then averaged for 
the final systematic uncertainty. Three types of systematic uncertainties 
are considered for this measurement. The first type is point-to-point 
uncorrelated uncertainties across $p_T$ bins. These include contributions 
from the choice of the invariant-mass fitting method for yield extraction 
(2\%--10\%) and the choice of mixed event background normalization 
(5\%--10\% for $p_T<2.25$~GeV/$c$). The second type of systematic 
uncertainties are correlated across $p_T$ bins. These include uncertainties 
due to the MPC energy scale (5\%-35\%, increasing with $p_T$ due to the 
convolution of the intrinsic 2\% energy-scale uncertainty with bin smearing 
from the physical $p_T$ spectrum), MB-trigger-efficiency (4\%), 
MPC-trigger-efficiency thresholds (5\%--30\%, decreasing in $p_T$), $\eta$ 
meson reconstruction efficiency (8\%), and high-energy cluster merging 
studied with the fully reconstructed single $\eta$ meson simulations 
(3\%--10\%, increasing in $p_T>3.5$~GeV/$c$). Finally, a global-scale 
uncertainty of 9.3\% comes from the Vernier-scan measurement needed for the 
integrated-luminosity determination. Correlated systematic uncertainties are 
treated as fully correlated across $p_T$ bins. The global-scale uncertainty 
is treated separately from the other correlated systematic uncertainties but 
is fully correlated between this forward measurement and the midrapidity 
measurement discussed in section~\ref{ssec:etaxs2}. The correlation 
matrix of the systematic uncertainties is included in 
HEPData~\cite{hepdata}.

\subsubsection{\label{sec:xsresults2}Cross-section results}

\begin{table*}[htb]
\caption{\label{tab:table1}The invariant cross section of inclusive 
forward $\eta$ mesons at $\sqrt{s}=500$ GeV, as shown in 
Fig.~\ref{fig:finalcrosssection}. A global 9.3\% scale systematic 
uncertainty is not included.}
\begin{ruledtabular}
\begin{tabular}{ccccccc}
  $p_T$ bin [GeV/$c$] & $p_T$ [GeV/$c$] &$E\frac{d^3\sigma}{dp^3}$ [mb GeV$^{-2}c^3$] & $\sigma_{\rm stat}$ & $\sigma_{\rm syst}$ (Uncorrelated) & $\sigma_{\rm syst}$ (Correlated) & $\sigma_{\rm syst}$ (Total)\\ \hline
   1.00 to 1.25 & 1.125 & 1.89 $\times$ 10$^{-1}$ & 2.42 $\times$ 10$^{-2}$ & 3.57 $\times$ 10$^{-2}$ & 1.94 $\times$ 10$^{-2}$ & 4.07 $\times$ 10$^{-2}$ \\ 
   1.25 to 1.50 & 1.375 & 5.48 $\times$ 10$^{-2}$ & 5.05 $\times$ 10$^{-3}$ & 7.76 $\times$ 10$^{-3}$ & 7.09 $\times$ 10$^{-3}$ & 1.05 $\times$ 10$^{-2}$ \\ 
   1.50 to 1.75 & 1.625 & 1.64 $\times$ 10$^{-2}$ & 1.79 $\times$ 10$^{-3}$ & 2.73 $\times$ 10$^{-3}$ & 2.16 $\times$ 10$^{-3}$ & 3.46 $\times$ 10$^{-3}$ \\ 
   1.75 to 2.00 & 1.875 & 8.88 $\times$ 10$^{-3}$ & 9.93 $\times$ 10$^{-4}$ & 1.34 $\times$ 10$^{-3}$ & 9.21 $\times$ 10$^{-4}$ & 1.63 $\times$ 10$^{-3}$ \\ 
   2.00 to 2.25 & 2.125 & 4.07 $\times$ 10$^{-3}$ & 3.70 $\times$ 10$^{-4}$ & 2.87 $\times$ 10$^{-4}$ & 5.14 $\times$ 10$^{-4}$ & 5.88 $\times$ 10$^{-4}$ \\ 
   2.25 to 2.50 & 2.375 & 2.04 $\times$ 10$^{-3}$ & 1.61 $\times$ 10$^{-4}$ & 2.00 $\times$ 10$^{-4}$ & 2.67 $\times$ 10$^{-4}$ & 3.34 $\times$ 10$^{-4}$ \\ 
   2.50 to 2.75 & 2.625 & 1.02 $\times$ 10$^{-3}$ & 8.57 $\times$ 10$^{-5}$ & 9.40 $\times$ 10$^{-5}$ & 1.34 $\times$ 10$^{-4}$ & 1.64 $\times$ 10$^{-4}$ \\ 
   2.75 to 3.00 & 2.875 & 4.98 $\times$ 10$^{-4}$ & 4.84 $\times$ 10$^{-5}$ & 2.97 $\times$ 10$^{-5}$ & 6.40 $\times$ 10$^{-5}$ & 7.05 $\times$ 10$^{-5}$ \\ 
   3.00 to 3.25 & 3.125 & 3.21 $\times$ 10$^{-4}$ & 3.78 $\times$ 10$^{-5}$ & 2.40 $\times$ 10$^{-5}$ & 5.06 $\times$ 10$^{-5}$ & 5.60 $\times$ 10$^{-5}$ \\ 
   3.25 to 3.50 & 3.375 & 1.77 $\times$ 10$^{-4}$ & 2.57 $\times$ 10$^{-5}$ & 1.05 $\times$ 10$^{-5}$ & 2.71 $\times$ 10$^{-5}$ & 2.91 $\times$ 10$^{-5}$ \\ 
   3.50 to 3.75 & 3.625 & 9.50 $\times$ 10$^{-5}$ & 5.06 $\times$ 10$^{-6}$ & 2.14 $\times$ 10$^{-6}$ & 3.29 $\times$ 10$^{-5}$ & 3.30 $\times$ 10$^{-5}$ \\ 
   3.75 to 4.00 & 3.875 & 4.96 $\times$ 10$^{-5}$ & 2.39 $\times$ 10$^{-6}$ & 2.77 $\times$ 10$^{-6}$ & 1.61 $\times$ 10$^{-5}$ & 1.64 $\times$ 10$^{-5}$ \\ 
   4.00 to 4.25 & 4.125 & 2.79 $\times$ 10$^{-5}$ & 1.30 $\times$ 10$^{-6}$ & 1.30 $\times$ 10$^{-6}$ & 8.58 $\times$ 10$^{-6}$ & 8.68 $\times$ 10$^{-6}$ \\ 
   4.25 to 4.50 & 4.375 & 1.68 $\times$ 10$^{-5}$ & 7.68 $\times$ 10$^{-7}$ & 8.86 $\times$ 10$^{-7}$ & 4.73 $\times$ 10$^{-6}$ & 4.81 $\times$ 10$^{-6}$ \\ 
   4.50 to 4.75 & 4.625 & 1.01 $\times$ 10$^{-5}$ & 4.92 $\times$ 10$^{-7}$ & 9.40 $\times$ 10$^{-7}$ & 3.00 $\times$ 10$^{-6}$ & 3.14 $\times$ 10$^{-6}$ \\ 
   4.75 to 5.00 & 4.875 & 6.89 $\times$ 10$^{-6}$ & 3.55 $\times$ 10$^{-7}$ & 1.13 $\times$ 10$^{-7}$ & 2.14 $\times$ 10$^{-6}$ & 2.14 $\times$ 10$^{-6}$ \\ 
   5.00 to 5.25 & 5.125 & 4.31 $\times$ 10$^{-6}$ & 2.29 $\times$ 10$^{-7}$ & 1.40 $\times$ 10$^{-7}$ & 1.40 $\times$ 10$^{-6}$ & 1.41 $\times$ 10$^{-6}$ \\ 
   5.25 to 5.50 & 5.375 & 2.73 $\times$ 10$^{-6}$ & 1.55 $\times$ 10$^{-7}$ & 1.10 $\times$ 10$^{-7}$ & 8.55 $\times$ 10$^{-7}$ & 8.62 $\times$ 10$^{-7}$ \\ 
   5.5 to 6.0 & 5.750 & 1.36 $\times$ 10$^{-6}$ & 7.66 $\times$ 10$^{-8}$ & 2.17 $\times$ 10$^{-7}$ & 4.82 $\times$ 10$^{-7}$ & 5.29 $\times$ 10$^{-7}$ \\ 
   6.0 to 6.5 & 6.250 & 6.20 $\times$ 10$^{-7}$ & 5.01 $\times$ 10$^{-8}$ & 6.62 $\times$ 10$^{-9}$ & 2.48 $\times$ 10$^{-7}$ & 2.49 $\times$ 10$^{-7}$ \\ 
\end{tabular}
\end{ruledtabular}
\end{table*}

\begin{figure*}[h!]
\includegraphics[width=0.99\linewidth]{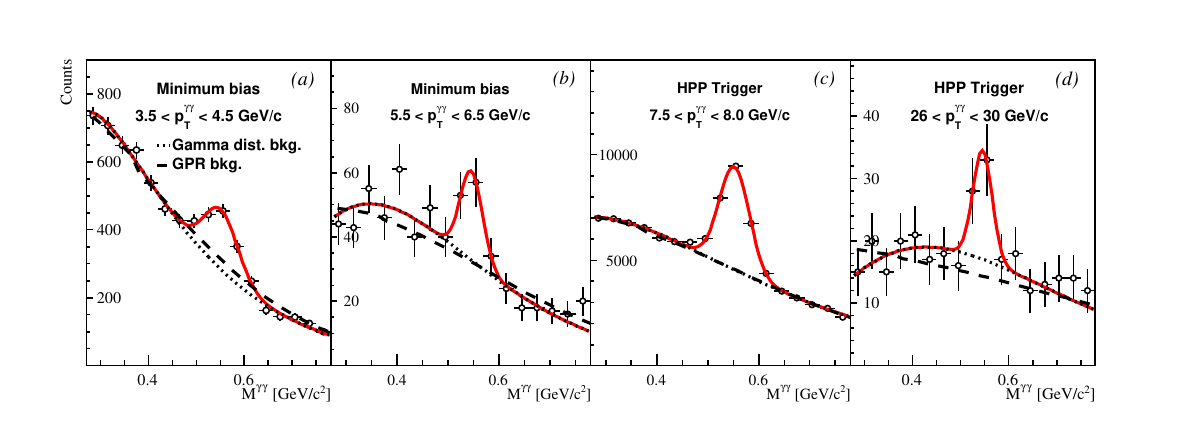}
\caption{\label{fig:invmass_midrap}Examples of two-photon invariant-mass 
distributions in the Central EMCal from the 2013 $\sqrt{s}=510$ GeV run in 
(a,b) MB and (c,d) HPP-triggered data. The black open circles represent the 
real diphoton cluster pairs while the red solid curves show the combined fit 
of the gamma-distribution background and signal while the gray dotted curves 
show only the gamma-distribution-background fit. The black dashed curves 
show the background description from Gaussian process regression.}
\end{figure*}

\begin{figure*}[htb]
\includegraphics[width=0.99\linewidth]{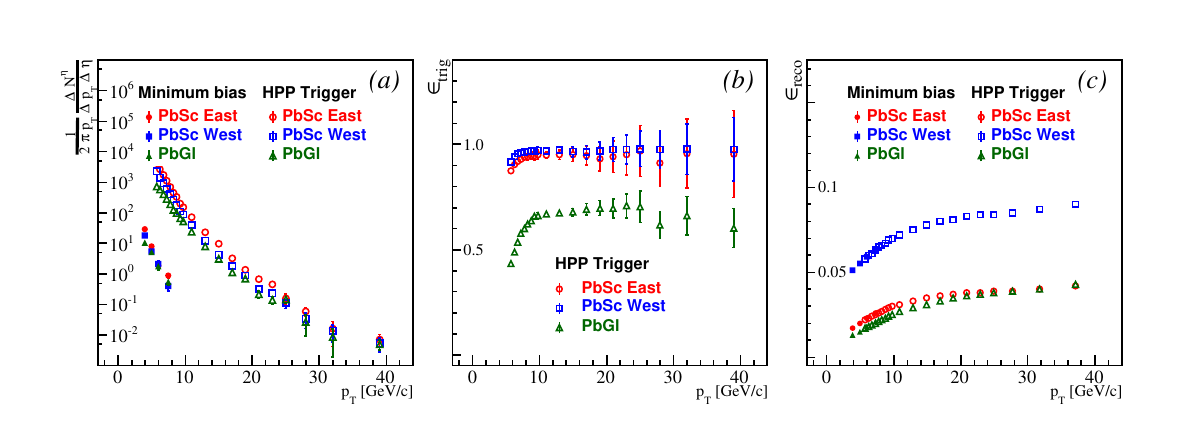}
\caption{\label{fig:yields_trigeff_recoeff}The inputs to the midrapidity 
$\eta$ meson cross section at $\sqrt{s}=510$ GeV for both the MB and 
HPP-triggered dataset at pseudorapidities $|\eta|<0.35$. (a) The raw yields 
as a function of $\eta$ meson $p_T$, (b) the trigger efficiency as a 
function of $\eta$ meson $p_T$, (c) the reconstruction efficiency as a 
function of $\eta$ meson $p_T$.}
\end{figure*}

The final invariant cross section of inclusive $\eta$ mesons for 
$1.0<p_T<6.5$~GeV/$c$ and $3.0<|\eta|<3.8$ is shown in 
Fig.~\ref{fig:finalcrosssection} and listed in Table~\ref{tab:table1}. The 
cross section was measured independently in the south and north MPCs and 
agrees within 4\%. The final cross section is taken as the weighted mean of 
the cross sections in each MPC arm. In the $p_T$ region where the MB and 
MPC-triggered datasets overlap ($3.5<p_T<4.5$ GeV/$c$), the cross sections 
are consistent within the statistical uncertainties and only the 
MPC-triggered cross section, being significantly more precise, is included 
in the final cross section. A final correction must be applied because the 
cross sections are reported at their bin centers in a steeply falling 
distribution. The correction follows a standard 
procedure~\cite{PhysRevC.81.064904,LAFFERTY1995541}, in which a smooth 
function is fit to the cross section and is then used to scale the cross 
section in each $p_T$ bin by the ratio of its average value in the bin to 
its value at the bin center. In this analysis, the MB and 
MPC-triggered cross sections were fit with separate exponential 
distributions which provided a good description of the data points. As the 
bin widths are relatively narrow, the corrections are correspondingly small, 
ranging from 1\% for the bins with widths of 0.25 GeV/$c$ to 4\% for the bins 
with widths of 0.5 GeV/$c$.

The cross section is compared to two NLO-pQCD calculations. The 
first uses the new ALMSS $\eta$-meson FFs~\cite{etaFFs} along with the 
MSHT20 PDFs~\cite{MSHT20}. The second~\cite{Vogelsang_priv} uses the old 
AESSS $\eta$ meson FFs~\cite{PhysRevD.83.034002} together with the CTEQ18 
PDFs~\cite{PhysRevD.103.014013}. In both cases, the calculations agree well 
with the data within statistical, systematic, and theoretical uncertainties 
in the perturbative regime above $p_T>2$~GeV/$c$. As this measurement was 
itself included in the ALMSS analysis, it is natural that the prediction 
using the ALMSS FFs matches nicely with the data. To test the actual 
predictive capability of the ALMSS FFs, the calculation was repeated with a 
modified FF fit that excluded the forward $\eta$-meson cross section 
reported here from the fit data. In this case, the predictions remain 
consistent with the measurement, indicating both that the ALMSS FFs are able 
to describe the forward $\eta$ meson cross section at $500$ GeV in a fully 
predictive manner and that the PHENIX measurement is consistent with other 
world data. An additional theoretical uncertainty is included in the ALMSS 
predictions related to the fact that the uncertainties on the ALMSS FFs were 
estimated using Monte-Carlo replicas rather than the Lagrange-multiplier 
method used in the AESSS FFs. The 1$\sigma$ standard deviation of the cross 
section predictions across the final set of ALMSS $\eta$ FF replicas are 
shown in Fig.~\ref{fig:finalcrosssection}. This uncertainty band represents 
the statistical dispersion of the fit ensemble rather than a fixed tolerance 
interval in $\chi^2$. The replica-based method more naturally reflects 
correlated uncertainties among fit parameters but should not be directly 
compared in magnitude to an uncertainty derived from the Lagrange-multiplier 
method. In either case, the replica uncertainty and the Lagrange multiplier 
uncertainty are quite small in comparison with the theoretical uncertainty 
from the renormalization scale.

\subsection{\label{ssec:etaxs2} The midrapidity $\eta$ meson cross 
section at $\sqrt{s}=510$~GeV}

In 2013, PHENIX collected a high luminosity dataset of proton 
collisions at $\sqrt{s}=510$ GeV. The cross section of $\eta$ mesons 
is measured independently in each arm and calorimeter type: PbSc west, 
PbSc east, and PbGl. It is also measured independently in events 
satisfying only the MB narrow event vertex ($|z|<10$ cm) 
condition and events that are required to have fired the HPP trigger 
with an additional offline vertex cut of 10 cm. The MB cross 
section is measured differentially in $p_T$ from 3.5 to 8.0~GeV/$c$. The 
HPP cross section is measured in $p_T$ from 5.5 to 44 GeV/$c$.

\subsubsection{\label{sssec:midrapetayields}Yields}

The invariant-mass distributions of photon cluster pairs in the central 
EMCal are formed in the same manner as Eq.~(\ref{eq:2}). The cluster-energy 
asymmetry requirement is loosened to 0.8 and the cluster separation cut that 
was present in the forward cross section is removed as cluster merging is 
negligible for $\eta$ mesons in the kinematic range measured. A ToF 
requirement of $|\textrm{ToF}|<10$~ns was applied to reduce the impact of 
out-of-time pileup events. As was done in the forward analysis, the raw 
$\eta$-meson yield for each $p_T$ bin is determined from a simultaneous 
signal and background fit described by a Gaussian signal and multiple 
choices for a description of a background. The background is found to be 
well described across the entire $p_T$ range by a gamma-distribution 
function, which is thus taken as nominal. Invariant-mass distributions of 
photon cluster pairs from several representative $p_T$ regions are shown in 
Fig.~\ref{fig:invmass_midrap}. The resulting yield distribution, shown in 
Fig.~\ref{fig:yields_trigeff_recoeff}(a), is robust to small variations in 
the invariant-mass fit region. A systematic uncertainty on the yield 
extraction is defined as the relative difference of the yield determined 
from the nominal background description and the yield determined with a
2nd-order-polynomial background fit and a GPR background fit with a 
radial-basis-function kernel.

\subsubsection{\label{sssec:mideffs}Efficiencies and Corrections}

The MB-trigger efficiency is determined by finding the likelihood that an 
event containing an $\eta$ meson reconstructed in the EMCal within a 
HPP-trigger sample also fired the MB trigger. It is found to be 
$0.91{\pm}0.01$, consistent with the value used in the forward cross section 
measurement in section~\ref{sssec:forwardeffs}. For the HPP trigger 
efficiency, a set of events uncorrelated with the $\eta$ meson photon 
clusters was obtained by requiring the opposite arm to have fired the 
trigger. Among these events, the fraction of $\eta$ mesons that had at least 
one of their underlying photon clusters also fire the HPP trigger defines 
the trigger efficiency. The efficiency reaches a plateau value of 0.95 at 
$p_T \gtrsim 8$ GeV/$c$ for the PbSc calorimeter and 0.67 at $p_T \gtrsim 
9.5$ GeV/$c$ for the PbGl, as seen in 
Fig.~\ref{fig:yields_trigeff_recoeff}(b). The difference in efficiencies 
between calorimeter types arises from a more-restrictive trigger masking in 
the case of the PbGl calorimeter. The efficiencies in the low $p_T$ turn-on 
region were found to be consistent with an independent method of determining 
the efficiency using MB events.

The reconstruction efficiency was calculated using a well-validated PHENIX 
fast-Monte-Carlo framework~\cite{PhysRevD.93.011501, PhysRevLett.130.251901} 
that contains a realistic description of the detector geometry with known 
smearing and inefficiency effects~\cite{APHECETCHE2003521}. A sample of 
single $\eta$ mesons was generated flat within $1<p_T<50$~GeV/$c$, weighted 
by the $\eta$ meson $p_T$ spectrum determined from a NLO-pQCD + AESSS 
calculation, and then fed into the detector simulation. The total 
reconstruction efficiency accounts for corrections stemming from acceptance, 
smearing due to imperfect detector resolution, and the 
cluster-energy-asymmetry cut, which is shown in 
Fig.~\ref{fig:yields_trigeff_recoeff}(c). In the acceptance of the 
west(east) EMCal, an additional correction is needed for the 28\%(8\%) 
chance that either of the $\eta$-meson decay photons convert to $e^+e^-$ in 
detector material near the beam pipe and are unable to be reconstructed in 
the EMCal.

\subsubsection{\label{sec:systs2}Systematic Uncertainties}

Systematic uncertainties for the midrapidity cross section are determined 
independently for the PbSc west, PbSc east, and PbGl, and then averaged for 
the final systematic uncertainty. The uncorrelated systematic uncertainties 
for the midrapidity cross section include the description of the 
combinatorial background used in the yield extraction (2\%--10\%), the 
description of detector resolution (1.5\%) and acceptance (3\%) in the Monte 
Carlo, the EMCal energy nonlinearity (1\%--10\% increasing in $p_T$), the 
HPP trigger efficiency in the turn-on region (6\%--3\% decreasing in $p_T$), 
the multiple collision correction (2\%), and the photon-conversion 
correction (4.3\%). Correlated uncertainties are due to the EMCal energy 
scale (4\%), and MB-trigger efficiency coupled with the normalization of the 
HPP and MB datasets due to trigger scale downs (1\%--2\%). As in the forward 
cross section, correlated systematic uncertainties are treated as fully 
correlated across $p_T$ bins and a 9.3\% global scale uncertainty is fully 
correlated with the forward measurement.

\subsubsection{\label{sec:xsresults}Cross section results}

The final invariant cross section of inclusive $\eta$ mesons for 
$3.5<p_T<44$ GeV/$c$ and $|\eta|<0.35$ is shown in 
Fig.~\ref{fig:finalcrosssectionmidrap} and listed in Table~\ref{tab:table2}. 
The cross section has been measured independently in the PbSc west, PbSc 
east, and PbGl calorimeters, which agree within the statistical and 
systematic uncertainties. The final cross section is calculated as the 
weighted average of these independent cross-section measurements and is 
reported at the mean-$p_T$ value in each bin, so the bin shift correction 
introduced in the forward cross section was not applied. In the $p_T$ 
overlap region of the MB and HPP-triggered data between 5.5 and 
8.0 GeV/$c$, the independently determined cross sections based on the two 
datasets agree with one another. Only the HPP-triggered cross section is 
included in the final cross section within this overlap region. 
Figure~\ref{fig:finalcrosssectionmidrap} shows the comparison of the cross 
section to NLO-pQCD calculations~\cite{etaFFs, Vogelsang_priv2025} with the 
ALMSS $\eta$ meson FFs and AESSS FFs, showing good agreement with both 
predictions within statistical, systematic, and theoretical uncertainties. 
As in the case of the forward cross section, the ALMSS FF fit was repeated 
excluding this specific measurement and the corresponding predicted cross 
section utilizing the modified FFs was found to be consistent with the 
prediction from the full ALMSS FFs.

\subsection{\label{ssec:etaxs3}The midrapidity $\eta/\pi^0$ ratio at 
$\sqrt{s}=510$ GeV}

\begin{figure*}[htb]
\begin{minipage}{0.48\linewidth}
\includegraphics[width=0.99\linewidth]{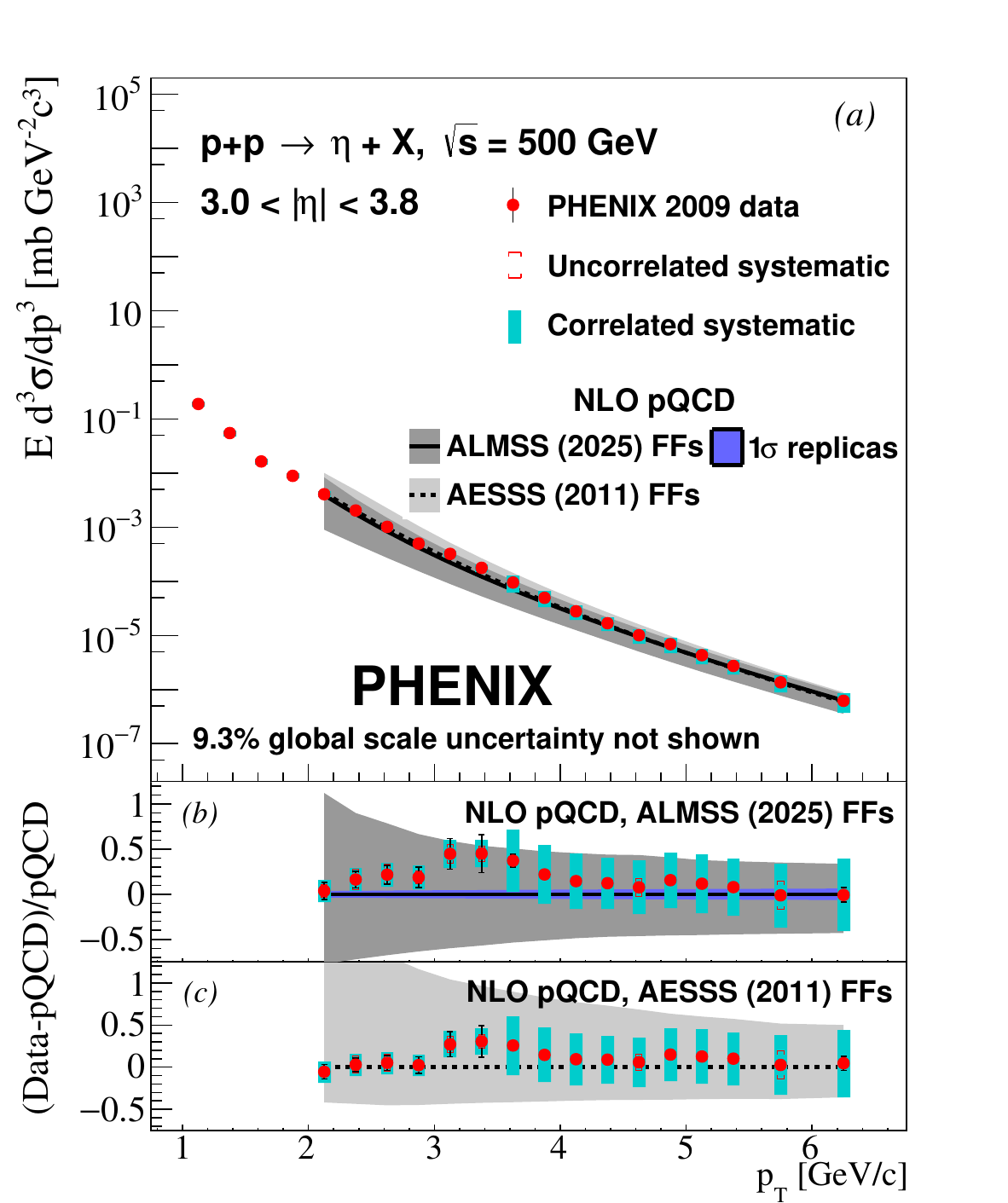}
\caption{\label{fig:finalcrosssectionmidrap}The invariant cross section 
of inclusive $\eta$ mesons at midrapidity in $\sqrt{s}=510$ GeV $p$$+$$p$ 
collisions. Statistical uncertainties are shown in black uncertainty bars, 
systematic uncertainties uncorrelated across $p_T$ are shown in red 
uncertainty bars, and systematic uncertainties correlated across $p_T$ are 
shown in blue uncertainty boxes. A global 9.3\% systematic uncertainty is not 
shown. NLO-pQCD predictions using the AESSS~(2011)~\cite{PhysRevD.83.034002} 
$\eta$~FFs and the ALMSS~(2025)~\cite{etaFFs} $\eta$~FFs are shown as the 
dashed and solid line, respectively, and their corresponding 
theoretical uncertainties are displayed as the gray uncertainty bands. The 
blue uncertainty band shows the 1$\sigma$ standard deviation of the cross 
section predictions across the final set of ALMSS $\eta$~FF replicas.}
\end{minipage}
\hspace{0.3cm}
\begin{minipage}{0.48\linewidth}
\includegraphics[width=0.99\linewidth]{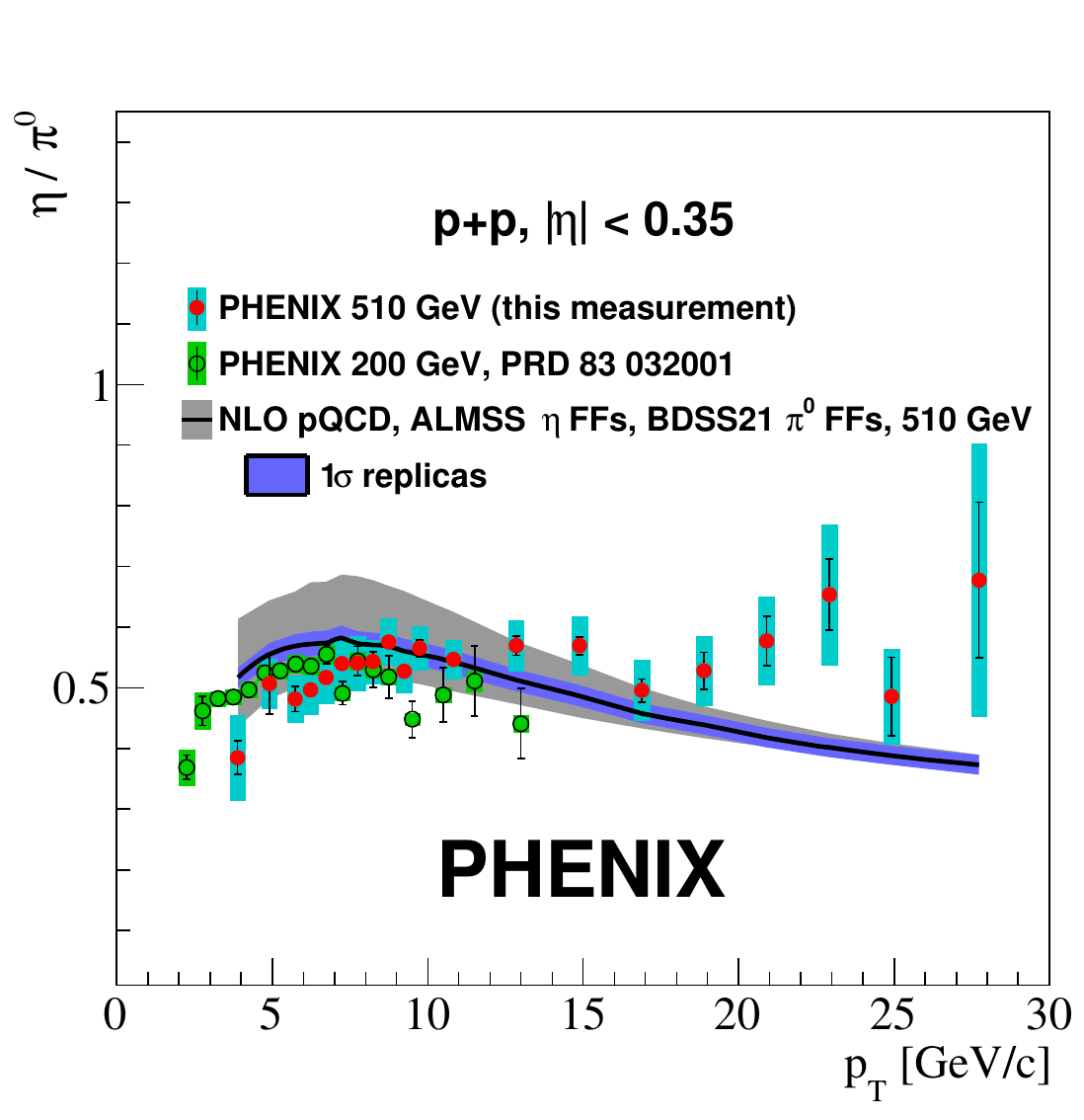}
\vspace{1.8cm}
\caption{\label{fig:finaletapi0rat}The $\eta/\pi^0$ ratio at 
midrapidity in $\sqrt{s}=510$ GeV $p$$+$$p$ collisions. Statistical 
uncertainties are shown in black uncertainty bars and systematic 
uncertainties in blue uncertainty boxes. NLO-pQCD predictions at 
$\sqrt{s}=510$ GeV using the ALMSS~(2025) $\eta$~FFs~\cite{etaFFs} 
and BDSS21 $\pi^0$~FFs~\cite{PhysRevD.105.L031502} are shown 
as the solid black line. The ratio measured by PHENIX at $\sqrt{s}=200$ 
GeV~\cite{PhysRevD.83.032001} is shown with the open black circles, 
where statistical uncertainties are represented by the black uncertainty bars 
and systematic uncertainties by the green uncertainty boxes. The theoretical 
scale uncertainty is represented by the gray uncertainty band while the blue 
uncertainty band shows the 1$\sigma$ standard deviation of the $\eta/\pi^0$ 
predictions across the final set of ALMSS $\eta$~FF replicas.}
\end{minipage}
\end{figure*}

Using the midrapidity $\pi^0$ cross section at $\sqrt{s}=510$ 
GeV~\cite{PhysRevD.93.011501} from the same 2013 dataset as the $\eta$ 
meson cross section, we report the measurement of the $\eta/\pi^0$ 
ratio with the highest recorded $p_T$ to date at RHIC. 
Figure~\ref{fig:finaletapi0rat} and Table~\ref{tab:table3} show the 
midrapidity $\eta/\pi^0$ ratio. The ratio can serve as a powerful input 
to FF global fits as any shared systematic uncertainties between the 
cross sections cancel.  However, as cautioned in~\cite{etaFFs}, care must be 
taken because inaccuracies in the individual $\eta$ and 
$\pi^0$ FFs can be obscured when looking at the ratio alone. 
Figure~\ref{fig:finaletapi0rat} shows the ratio along with an earlier 
PHENIX measurement of the midrapidity ratio at $\sqrt{s}=200$ 
GeV~\cite{PhysRevD.83.032001} compared to a NLO-pQCD calculation at 
$\sqrt{s}=510$ GeV that uses the ALMSS $\eta$ FFs and the BDSS21 
$\pi^0$ FFs~\cite{PhysRevD.105.L031502}.

At $p_T>8$~GeV/$c$, the ratio saturates at $0.54{\pm}0.03$, consistent with 
the behavior observed at lower energy from ISR~\cite{KOURKOUMELIS1979277}, 
PHENIX~\cite{PhysRevD.83.032001} and 
STAR~\cite{PhysRevC.81.064904,PhysRevD.86.051101} and higher energy results from 
ALICE~\cite{2018EPJC...78..263A}. This saturation of the ratio at high $p_T$ 
has been found across a wide range of additional measurements with different 
energies and collision systems~\cite{PhysRevC.104.054902}. This suggests 
that $\eta$ and $\pi^0$ production at high $p_T$ is generated through some 
universal fragmentation mechanism that is relatively robust to differences 
in energy, $p_T$, and the nuclear medium. Any difference between $\eta$ and 
$\pi^0$ production at high $p_T$ originating from the enhanced contribution 
of strange quark and gluon fragmentation into $\eta$ mesons appears to be 
limited. The high $p_T$ saturation behavior is consistent with empirical 
models of simple transverse-mass scaling of the $\pi^0$ spectrum. Recent 
ALICE results~\cite{2018EPJC...78..263A} have shown the breakdown of 
transverse mass scaling at low $p_T$, suggesting a more complicated 
interplay of fragmentation contributions at $p_T$ below the reach of this 
measurement.

\begin{table*}[htb]
\begin{minipage}{0.99\linewidth}
\caption{\label{tab:table2}The invariant cross section of inclusive 
midrapidity $\eta$ mesons at $\sqrt{s}=510$ GeV, as shown in 
Fig.~\protect\ref{fig:finalcrosssectionmidrap}. A global 9.3\% scale 
systematic uncertainty is not included.}
\begin{ruledtabular}
\begin{tabular}{ccccccc}
  $p_T$ bin [GeV/$c$] & $p_T$ [GeV/$c$] &$E\frac{d^3\sigma}{dp^3}$ [mb GeV$^{-2}c^3$] & $\sigma_{\rm stat}$ & $\sigma_{\rm syst}$ (Uncorrelated) & $\sigma_{\rm syst}$ (Correlated) & $\sigma_{\rm syst}$ (Total)\\ \hline
  3.5 to 4.5 & 3.89 & 2.827 $\times$ 10$^{-4}$ & 2.005 $\times$ 10$^{-5}$ & 5.297 $\times$ 10$^{-5}$ & 1.138 $\times$ 10$^{-5}$ & 5.42 $\times$ 10$^{-5}$ \\
  4.5 to 5.5 & 4.91 & 7.912 $\times$ 10$^{-5}$ & 7.584 $\times$ 10$^{-6}$ & 6.907 $\times$ 10$^{-6}$ & 3.320 $\times$ 10$^{-6}$ & 7.66 $\times$ 10$^{-6}$ \\
  5.5 to 6.0 & 5.73 & 2.345 $\times$ 10$^{-5}$ & 1.774 $\times$ 10$^{-7}$ & 2.489 $\times$ 10$^{-6}$ & 1.062 $\times$ 10$^{-6}$ & 2.71 $\times$ 10$^{-6}$ \\
  6.0 to 6.5 & 6.23 & 1.375 $\times$ 10$^{-5}$ & 1.181 $\times$ 10$^{-7}$ & 1.323 $\times$ 10$^{-6}$ & 6.273 $\times$ 10$^{-7}$ & 1.46 $\times$ 10$^{-6}$ \\
  6.5 to 7.0 & 6.73 & 8.283 $\times$ 10$^{-6}$ & 9.354 $\times$ 10$^{-8}$ & 8.149 $\times$ 10$^{-7}$ & 3.801 $\times$ 10$^{-7}$ & 8.99 $\times$ 10$^{-7}$ \\
  7.0 to 7.5 & 7.23 & 5.122 $\times$ 10$^{-6}$ & 5.987 $\times$ 10$^{-8}$ & 4.630 $\times$ 10$^{-7}$ & 2.361 $\times$ 10$^{-7}$ & 5.18 $\times$ 10$^{-7}$ \\
  7.5 to 8.0 & 7.74 & 3.109 $\times$ 10$^{-6}$ & 4.378 $\times$ 10$^{-8}$ & 2.994 $\times$ 10$^{-7}$ & 1.438 $\times$ 10$^{-7}$ & 3.33 $\times$ 10$^{-7}$ \\
  8.0 to 8.5 & 8.24 & 1.981 $\times$ 10$^{-6}$ & 3.215 $\times$ 10$^{-8}$ & 1.667 $\times$ 10$^{-7}$ & 9.193 $\times$ 10$^{-8}$ & 1.92 $\times$ 10$^{-7}$ \\
  8.5 to 9.0 & 8.74 & 1.351 $\times$ 10$^{-6}$ & 2.466 $\times$ 10$^{-8}$ & 1.163 $\times$ 10$^{-7}$ & 6.288 $\times$ 10$^{-8}$ & 1.32 $\times$ 10$^{-7}$ \\
  9.0 to 9.5 & 9.24 & 8.056 $\times$ 10$^{-7}$ & 1.859 $\times$ 10$^{-8}$ & 7.069 $\times$ 10$^{-8}$ & 3.757 $\times$ 10$^{-8}$ & 7.98 $\times$ 10$^{-8}$ \\
  9.5 to 10.0 & 9.74 & 6.064 $\times$ 10$^{-7}$ & 1.456 $\times$ 10$^{-8}$ & 5.374 $\times$ 10$^{-8}$ & 2.833 $\times$ 10$^{-8}$ & 6.09 $\times$ 10$^{-8}$ \\
  10 to 12 & 10.82 & 2.724 $\times$ 10$^{-7}$ & 4.618 $\times$ 10$^{-9}$ & 2.433 $\times$ 10$^{-8}$ & 1.277 $\times$ 10$^{-8}$ & 2.73 $\times$ 10$^{-8}$ \\
  12 to 14 & 12.85 & 8.120 $\times$ 10$^{-8}$ & 2.161 $\times$ 10$^{-9}$ & 8.427 $\times$ 10$^{-9}$ & 3.820 $\times$ 10$^{-9}$ & 9.28 $\times$ 10$^{-9}$ \\
  14 to 16 & 14.88 & 2.801 $\times$ 10$^{-8}$ & 7.052 $\times$ 10$^{-10}$ & 3.275 $\times$ 10$^{-9}$ & 1.321 $\times$ 10$^{-9}$ & 3.53 $\times$ 10$^{-9}$ \\
  16 to 18 & 16.88 & 9.913 $\times$ 10$^{-9}$ & 3.704 $\times$ 10$^{-10}$ & 1.278 $\times$ 10$^{-9}$ & 4.683 $\times$ 10$^{-10}$ & 1.36 $\times$ 10$^{-9}$ \\
  18 to 20 & 18.89 & 4.569 $\times$ 10$^{-9}$ & 2.555 $\times$ 10$^{-10}$ & 6.198 $\times$ 10$^{-10}$ & 2.161 $\times$ 10$^{-10}$ & 6.57 $\times$ 10$^{-10}$ \\
  20 to 22 & 20.91 & 2.157 $\times$ 10$^{-9}$ & 1.416 $\times$ 10$^{-10}$ & 3.203 $\times$ 10$^{-10}$ & 1.021 $\times$ 10$^{-10}$ & 3.36 $\times$ 10$^{-10}$ \\
  22 to 24 & 22.92 & 1.183 $\times$ 10$^{-9}$ & 9.528 $\times$ 10$^{-11}$ & 2.224 $\times$ 10$^{-10}$ & 5.604 $\times$ 10$^{-11}$ & 2.29 $\times$ 10$^{-10}$ \\
  24 to 26 & 24.92 & 5.264 $\times$ 10$^{-10}$ & 5.920 $\times$ 10$^{-11}$ & 7.977 $\times$ 10$^{-11}$ & 2.494 $\times$ 10$^{-11}$ & 8.36 $\times$ 10$^{-11}$ \\
  26 to 30 & 27.74 & 2.427 $\times$ 10$^{-10}$ & 3.201 $\times$ 10$^{-11}$ & 6.191 $\times$ 10$^{-11}$ & 1.151 $\times$ 10$^{-11}$ & 6.30 $\times$ 10$^{-11}$ \\
  30 to 34 & 31.77 & 5.068 $\times$ 10$^{-11}$ & 1.326 $\times$ 10$^{-11}$ & 1.685 $\times$ 10$^{-11}$ & 2.404 $\times$ 10$^{-12}$ & 1.70 $\times$ 10$^{-11}$ \\
  34 to 44 & 37.15 & 2.263 $\times$ 10$^{-11}$ & 4.594 $\times$ 10$^{-12}$ & 4.757 $\times$ 10$^{-12}$ & 1.074 $\times$ 10$^{-12}$ & 4.88 $\times$ 10$^{-12}$ \\
\end{tabular}
\end{ruledtabular}
\end{minipage}
\begin{minipage}{0.99\linewidth}
\caption{\label{tab:table3}The midrapidity $\eta/\pi^0$ at 
$\sqrt{s}=510$~GeV, as shown in Fig.~\protect\ref{fig:finaletapi0rat}.}
\begin{ruledtabular}
\begin{tabular}{cccccc}
  $p_T$ bin [GeV/$c$] & $p_T$ [GeV/$c$] & $\eta/\pi^0$ & $\sigma_{\rm stat}$ & $\sigma_{\rm syst}$ (Uncorrelated) \\ \hline
    3.5 to 4.5 & 3.89 & 3.851 $\times$ 10$^{-1}$ & 2.791 $\times$ 10$^{-2}$ & 7.027 $\times$ 10$^{-2}$ \\
    4.5 to 5.5 & 4.91 & 5.068 $\times$ 10$^{-1}$ & 5.040 $\times$ 10$^{-2}$ & 4.054 $\times$ 10$^{-2}$ \\ 
    5.5 to 6.0 & 5.73 & 4.816 $\times$ 10$^{-1}$ & 2.086 $\times$ 10$^{-2}$ & 3.846 $\times$ 10$^{-2}$ \\ 
    6.0 to 6.5 & 6.23 & 4.967 $\times$ 10$^{-1}$ & 4.398 $\times$ 10$^{-3}$ & 3.997 $\times$ 10$^{-2}$ \\ 
    6.5 to 7.0 & 6.73 & 5.167 $\times$ 10$^{-1}$ & 5.986 $\times$ 10$^{-3}$ & 4.319 $\times$ 10$^{-2}$ \\ 
    7.0 to 7.5 & 7.23 & 5.395 $\times$ 10$^{-1}$ & 6.522 $\times$ 10$^{-3}$ & 4.272 $\times$ 10$^{-2}$ \\ 
    7.5 to 8.0 & 7.74 & 5.407 $\times$ 10$^{-1}$ & 7.867 $\times$ 10$^{-3}$ & 4.454 $\times$ 10$^{-2}$ \\ 
    8.0 to 8.5 & 8.24 & 5.431 $\times$ 10$^{-1}$ & 9.122 $\times$ 10$^{-3}$ & 3.627 $\times$ 10$^{-2}$ \\ 
    8.5 to 9.0 & 8.74 & 5.755 $\times$ 10$^{-1}$ & 1.091 $\times$ 10$^{-2}$ & 3.841 $\times$ 10$^{-2}$ \\ 
    9.0 to 9.5 & 9.24 & 5.276 $\times$ 10$^{-1}$ & 1.259 $\times$ 10$^{-2}$ & 3.500 $\times$ 10$^{-2}$ \\ 
    9.5 to 10.0 & 9.74 & 5.651 $\times$ 10$^{-1}$ & 1.412 $\times$ 10$^{-2}$ & 3.635 $\times$ 10$^{-2}$ \\ 
    10 to 12 & 10.82 & 5.465 $\times$ 10$^{-1}$ & 9.602 $\times$ 10$^{-3}$ & 3.122 $\times$ 10$^{-2}$ \\ 
    12 to 14 & 12.85 & 5.694 $\times$ 10$^{-1}$ & 1.587 $\times$ 10$^{-2}$ & 4.153 $\times$ 10$^{-2}$ \\ 
    14 to 16 & 14.88 & 5.693 $\times$ 10$^{-1}$ & 1.486 $\times$ 10$^{-2}$ & 4.738 $\times$ 10$^{-2}$ \\ 
    16 to 18 & 16.88 & 4.956 $\times$ 10$^{-1}$ & 1.917 $\times$ 10$^{-2}$ & 4.985 $\times$ 10$^{-2}$ \\ 
    18 to 20 & 18.89 & 5.278 $\times$ 10$^{-1}$ & 3.064 $\times$ 10$^{-2}$ & 5.735 $\times$ 10$^{-2}$ \\ 
    20 to 22 & 20.91 & 5.774 $\times$ 10$^{-1}$ & 4.054 $\times$ 10$^{-2}$ & 7.302 $\times$ 10$^{-2}$ \\ 
    22 to 24 & 22.92 & 6.538 $\times$ 10$^{-1}$ & 5.836 $\times$ 10$^{-2}$ & 1.155 $\times$ 10$^{-1}$ \\ 
    24 to 26 & 24.92 & 4.856 $\times$ 10$^{-1}$ & 6.449 $\times$ 10$^{-2}$ & 7.901 $\times$ 10$^{-2}$ \\ 
    26 to 30 & 27.74 & 6.776 $\times$ 10$^{-1}$ & 1.285 $\times$ 10$^{-1}$ & 2.250 $\times$ 10$^{-1}$ \\ 
\end{tabular}
\end{ruledtabular}
\end{minipage}
\end{table*}

\section{\label{sec:summary}Summary}

The forward $\eta$-meson cross section at $\sqrt{s}=500$ GeV, as well 
as the midrapidity $\eta$-meson cross section and $\eta/\pi^0$ ratio at 
$\sqrt{s}=510$ GeV, were measured at PHENIX. Utilizing the cross 
sections, a new global $\eta$-meson-FF fit reported in 
Ref.~\cite{etaFFs} has been performed at NLO accuracy, with improved 
constraints compared to the previous $\eta$-meson 
FFs~\cite{PhysRevD.83.034002} and the ability to simultaneously 
describe $\eta$-meson production from RHIC, the LHC, 
and high-statistics $e^+e^-$ data from Belle. Theoretical predictions 
at NLO, utilizing both the updated FFs as well as the previous set of 
FFs, are consistent with the cross section measurements reported in 
this article within theoretical uncertainties.
\\ \\
\noindent {\textit Data Availability}: The data that support the findings of this 
article are not publicly available. The values in the plots and tables 
associated with this article are stored in HEPData~\cite{hepdata}.


\clearpage

\section*{ACKNOWLEDGMENTS}


We thank the staff of the Collider-Accelerator and Physics
Departments at Brookhaven National Laboratory and the staff of
the other PHENIX participating institutions for their vital
contributions.  
We also thank R. T. Martinez, R. Sassot, M. Stratmann, 
and W. Vogelsang for fruitful discussions.
We acknowledge support from the Office of Nuclear Physics in the
Office of Science of the Department of Energy,
the National Science Foundation,
Abilene Christian University Research Council,
Research Foundation of SUNY, and
Dean of the College of Arts and Sciences, Vanderbilt University
(U.S.A),
Ministry of Education, Culture, Sports, Science, and Technology
and the Japan Society for the Promotion of Science (Japan),
Conselho Nacional de Desenvolvimento Cient\'{\i}fico e
Tecnol{\'o}gico and Funda\c c{\~a}o de Amparo {\`a} Pesquisa do
Estado de S{\~a}o Paulo (Brazil),
Natural Science Foundation of China (People's Republic of China),
Croatian Science Foundation and
Ministry of Science and Education (Croatia),
Ministry of Education, Youth and Sports (Czech Republic),
Centre National de la Recherche Scientifique, Commissariat
{\`a} l'{\'E}nergie Atomique, and Institut National de Physique
Nucl{\'e}aire et de Physique des Particules (France),
J. Bolyai Research Scholarship, EFOP, HUN-REN ATOMKI, NKFIH,
MATE KKF, and OTKA (Hungary), 
Department of Atomic Energy and Department of Science and Technology (India),
Israel Science Foundation (Israel),
Basic Science Research and SRC(CENuM) Programs through NRF
funded by the Ministry of Education and the Ministry of
Science and ICT (Korea),
Ministry of Education and Science, Russian Academy of Sciences,
Federal Agency of Atomic Energy (Russia),
VR and Wallenberg Foundation (Sweden),
University of Zambia, the Government of the Republic of Zambia (Zambia),
the U.S. Civilian Research and Development Foundation for the
Independent States of the Former Soviet Union,
the Hungarian American Enterprise Scholarship Fund,
the US-Hungarian Fulbright Foundation,
and the US-Israel Binational Science Foundation.



%
 
\end{document}